\documentclass[fleqn,usenatbib]{mnras}

\usepackage{newtxtext,newtxmath}

\usepackage[T1]{fontenc}
\usepackage{ae,aecompl}

\usepackage{graphicx}	
\usepackage{amsmath}	
\usepackage{amssymb}	

\def\aj{AJ}
\def\apj{ApJ}
\def\apjl{ApJL}
\def\apjs{ApJ Suppl. Ser.}

\def\aap{A\&A}

\def\araa{ARAA}
\def\mnras{MNRAS}

\def\pasj{PASJ}

\def\sovast{Sov. Astron.}

\def\beq#1{\begin{equation}\label{#1}}
\def\eeq{\end{equation}}
\def\beqa#1{\begin{eqnarray}\label{#1}}
\def\eeqa{\end{eqnarray}}

\def\Eq#1{Eq.~(\ref{#1})} 
\def\eqn#1{~(\ref{#1})}
\def\myfrac#1#2{\left(\frac{#1}{#2}\right)}

\def\comment#1{\relax}

\title[SS433 masses]{On masses of the components in SS433}

\author[A.M. Cherepashchuk et al.]{
A.M. Cherepashchuk$^{1}$\thanks{E-mail: cherepashchuk@gmail.com},
K.A. Postnov$^{1}$\thanks{E-mail: pk@sai.msu.ru}
and 
A.A. Belinski$^1$\thanks{E-mail: aleks.sai@gmail.com}
\\
$^{1}$Sternberg Astronomical Institute, M.V. Lomonosov Moscow State University, 13, Universitetskij pr., 119234, Moscow, Russia\\
}

\date{Accepted XXX. Received YYY; in original form ZZZ}

\pubyear{2018}

\begin{document}
\label{firstpage}
\pagerange{\pageref{firstpage}--\pageref{lastpage}}
\maketitle

\begin{abstract}
A huge optical luminosity of the supercritical accretion disc and powerful stellar wind in the high-mass X-ray binary SS433 make it difficult to reliably estimate the mass ratio of the binary components from spectroscopic observations. We analyze different indirect methods of the mass ratio estimate. We show that with an account of the possible Roche lobe overflow by the optical star, the analysis of X-ray eclipses in the standard and hard X-ray bands suggests the estimate $q=M_\mathrm{x}/M_\mathrm{v}\gtrsim 0.3$. We argue that the double-peak hydrogen Brackett lines in SS433 should form not in the accretion disc but in a circumbinary envelope, suggesting a total mass of $M_\mathrm{v}+M_\mathrm{x}\gtrsim 40 M_\odot$. 
The observed long-term stability of the orbital period in SS433 $|\dot P_b/P_b|\le 1.793\times 10^{-14}$~s$^{-1}$ over $\sim 28$ year period is used to place an independent constraint of $q\gtrsim 0.6$ in SS433, confirming its being a Galactic microquasar hosting a superaccreting black hole.   
\end{abstract}




\section{Introduction}

The unique Galactic microquasar SS433 is a high-mass eclipsing X-ray binary with an orbital period of $P_b\simeq 13^\mathrm{d}.08$ at an advanced evolutionary stage with a precessing ($P_\mathrm{prec}\simeq 162^\mathrm{d}.3$) supercritical optically bright accretion disc and relativistic jets ($v_j/c\simeq 0.26$) \citep{1979ApJ...233L..63M,1981MNRAS.194..761C,1984ARA&A..22..507M,1989ASPRv...7..185C}. Over40 years of studies in the optical, X-ray and radio, many features of SS433 have been studied in detail (see the review \cite{2004ASPRv..12....1F} and references therein). However, the masses of the binary components remain to be a highly debatable issue. Meanwhile, the component masses in SS433 are important to understand the evolutionary stage of this peculiar X-ray binary. The point is that SS433 provides a unique example of a high-mass X-ray binary at the stage of the secondary mass exchange when the more massive optical component overfills its Roche lobe avoiding the formation of a common envelope and the system remains to be a semi-detached binary (the formation of a common envelope at this stage
is predicted by the modern theory of massive binary evolution, see e.g. \cite{MT1988}). At this stage, the removal of the mass and angular momentum from the system is mediated by the supercritical accretion disc around the relativistic component with the formation of powerful wind outflow from the disc ($v_w\simeq 2000$~km~s$^{-1}$) and relativistic jets. 

This peculiar feature of SS433 has recently been addressed by \cite{2017MNRAS.471.4256V}. These authors noted that when the donor star in a high-mass X-ray binary has a radiative envelope and starts filling the Roche lobe and the mass ratio of the binary is $q=M_\mathrm{x}/M_\mathrm{v}\gtrsim 0.29$ (here $M_\mathrm{x}$ and $M_\mathrm{v}$ is the mass of the relativistic accretor and visual donor star, respectively), the formation of the common envelope can be avoided. The system evolves as a semi-detached binary with rapid but stable Roche lobe overflow when the mass transfer through the inner Lagrangian point $L_1$ is expelled from the vicinity of the relativistic star (the so-called isotropic re-emission mode, or 'SS433-like mode' of mass loss from the system). If the mass ratio $q\lesssim 0.29$, the high-mass X-ray binary at the secondary mass exchange stage  inevitably passes through the common envelope stage, and in this case depending on the  initial angular momentum either a close WR+BH (Cyg X-3 - like) binary forms or a Thorne-Zytkow object is produced \citep{1977ApJ...212..832T}. The authors \citep{2017MNRAS.471.4256V} stress that in the case of a neutron star, the mass ratio $q$ is much smaller than 0.29 and the binary system always evolves through the common envelope stage with the most likely formation of a Thorne-Zytkow object, unless the orbital period is longer than about 100 days.

Due to the huge optical luminosity of the supercritical accretion disc and powerful stellar wind in SS433, absorption lines in the optical companion spectrum are difficult to recognize, and the emission lines related to the precessing accretion disc are blended with emission lines from the powerful radial stellar wind. This makes it impossible to reliably estimate the dynamical mass ratio of the components in SS433. 

Of the recent spectroscopic mass estimates in SS433 we note the paper by \cite{2008ApJ...676L..37H} ($m_x=4.3\pm 0.8 M_\odot$, $m_o=12.3\pm 3.3. M_\odot$) and by \cite{2010ApJ...709.1374K} ($m_x=2.5^{+0.7}_{-0.6} M_\odot$, $m_o=10.4^{+2.3}_{-1.9}  M_\odot$), with the last study not excluding a neutron star in SS433. With all accuracy of spectroscopic observations carried out in these papers, the reliability of the mass estimates in SS433 can be put in doubt. For example,  \cite{2010A&A...521A..81B,2011A&A...531A.107B,2011A&A...534A.112B} showed that the absorption lines in the SS433 spectra that were used to estimate the component masses can also be formed not in the optical star photosphere but in a circumbinary shell whose inner region rotates with a velocity of $\sim 250$~km s$^{-1}$. This shell can also be found in the radio, where an image of the equatorial outflow perpendicular to relativistic jets was found \citep{2001ApJ...562L..79B}.
Two-peaked emission hydrogen lines found by 
IR \citep{1988AJ.....96..242F} and optical
\citep{2010MNRAS.408....2P} spectroscopic observations also suggest the presence of a circumbinary gas shell. 

In this paper, we analyze different possibilities to determine the component mass ratio in SS433 and discuss the reliability of the estimates of the mass ratio and total mass. In particular, we derive a new constraint on the component mass ratio from observational limits on the orbital period change with taking into account the possible mass-loss through the outer Lagrangian point L$_2$.

\section{The mass ratio estimate from the analysis of X-ray eclipses}

In the standard X-ray 2-10 keV range, the width of the primary X-ray eclipse of SS433 is rather high, $\sim 2^\mathrm{d}.4$ \citep{1989PASJ...41..491K,1989A&A...218L..13B,1996PASJ...48..619K}, with an orbital period of $\sim 13^\mathrm{d}.08$. In this case, the model of the eclipse of a thin relativistic jet by the optical star filling its Roche lobe yields a small mass ratio $q\simeq 0.15$ \citep{1989PASJ...41..491K,1989A&A...218L..13B,1996PASJ...48..619K}. This is significantly smaller than the critical mass ratio $q=0.29$ \citep{2017MNRAS.471.4256V}, and in this case, it is not clear why SS433 does not enter the common envelope stage and remains a semi-detached binary. 

Recently, studies on the stability of mass transfer in massive close binaries appeared \citep{2015MNRAS.449.4415P,2017MNRAS.465.2092P}. It is shown that nozzle-like mass transfer through the inner Lagrangian point $L_1$ is restricted by gas-dynamic and thermodynamic effects, and the optical star can steadily overfill its Roche lobe and even expel the matter through the outer Lagrangian point $L_2$. Thus, in the analysis of the 2-10 keV X-ray eclipses, it should be taken into account that the optical star radius can exceed the mean Roche lobe radius. This suggests that the mass ratio estimate $q=0.15$ obtained in \citep{1989PASJ...41..491K,1989A&A...218L..13B,1996PASJ...48..619K} is only a lower limit. Due to a weak dependence of the mean Roche lobe radius on the mass ratio ($R_{RL}/a\sim 0.38 q^{0.208}$, where $a$ is the binary orbital separation), a $\sim 15\%$ increase in the optical star radius increases the mass ratio estimate by a factor of two. Thus, a $\sim 10-20\%$ Roche lobe overflow by the optical star enables the X-ray eclipse width in SS433  to be described for $q\simeq 0.3$. 

The analysis of hard X-ray eclipses in SS433 (20-60 keV) observed by INTEGRAL with an account for the precession flux variability suggests the component mass ratio $q\gtrsim 0.3$ \citep{2013MNRAS.436.2004C}. In this analysis, the optical star was assumed to fill its Roche lobe and to eclipse an extended quasi-isothermal hot corona above the accretion disc. In addition, precession variability of the hard X-ray flux from SS433 limiting the corona height above the accretion disc was taken into account. Clearly, in this model, an account for a significant overfilling of the Roche lobe by the optical star will enhance the inequality $q>0.3$.

Thus, the possibility of the Roche lobe overflow by the optical star in the analysis of standard and hard X-ray eclipses allows us to estimate the component mass ratio in SS433 $q>0.3$, which exceeds the critical mass ratio $q=0.29$ for the stable evolution of the binary system without the common envelope formation \citep{2017MNRAS.471.4256V}.

\section{The estimate of the total mass of SS433 from the analysis of circumbinary envelope}

The analysis of the stationary H$_\alpha$ emission in the SS433 spectrum carried out in \cite{2001ApJ...562L..79B,2010A&A...521A..81B,2011A&A...531A.107B,2011A&A...534A.112B} and \cite{2010MNRAS.408....2P} by the line decomposition in the Gaussian components suggests the presence of a circumbinary shell rotating with a velocity of $\sim 250$~km s$^{-1}$ at the inner radius. These authors concluded that the total mass of the binary system in SS433 exceeds $\sim 40 M_\odot$ and the relativistic object is undoubtedly a high-mass black hole. The spectroscopic appearances of such a shell are also consistent with radio observations that reveal traces of the equatorial gas outflow from SS433 in the plane perpendicular to the relativistic jets \citep{2001ApJ...562L..79B}. 

The double-peak stationary hydrogen lines were also detected in the IR spectroscopic observations by  \cite{1988AJ.....96..242F} (Paschen series) and \cite{2017ApJ...841...79R} (Brackett series). \cite{1988AJ.....96..242F} interpret the double-peak hydrogen lines by two models: by the model of a circumbinary shell (suggesting a high total mass of SS433) and by the model in which these lines are produced in the accretion disc (and then the relativistic object can be a neutron star with a mass of $\lesssim 1.4 M_\odot$). 

On the other hand, \cite{2017ApJ...841...79R} argue that the double-peak stationary hydrogen emissions are formed in the accretion disc around a 1.4 $M_\odot$ neutron star. Their arguments are based on high-resolution IR spectroscopic observations revealing broad line wings that can be due to, as these authors believe, the widening by a rapid rotation in the inner parts of the accretion disc. \cite{2017ApJ...841...79R} carried out a numerical simulation of the emission line profiles from the accretion disc around a neutron star and obtained good agreement of theoretical profiles (double-peak profile with broad wings) with the IR spectroscopic observations of the Brackett hydrogen lines in SS433.

However, \cite{2017ApJ...841...79R} have ignored an important observational fact: in the middle of X-ray eclipses in SS433, the centre of the accretion disc should be screened by the optical star. As the orbital inclination in SS433 is reliably known from the analysis of moving emission limes ($i\simeq 79^\circ$), in the case of a circular orbit at the moment of the eclipse the distance between the accretion disc center and the optical star center is $\Delta_{min}=\cos i$ (in units of the orbital separation $a$). For $i=79^\circ$ we have $\Delta_{min}\approx 0.191$, which is definitely smaller than the radius of the optical star that fills or even overfills its Roche lobe. Therefore, in the case of a circular orbit, in SS433 a significant part of the central part of the accretion disc around the relativistic object in the middle of the X-ray eclipse should be screened by the optical star. Therefore, the broad wings of the emission lines, if formed in the rapidly rotating central parts of the accretion disc, should disappear at the middle of the X-ray eclipse, which is not observed \citep{2017ApJ...841...79R}. Additionally, the broad wings of the lines from the disc would strongly vary with the phase of the 162.3-day precession period.

In the case of an elliptical orbit, $\Delta_{min}$ can be larger than $\cos i$ at the middle of the X-ray eclipse, and the central parts of the accretion disc can remain non-eclipsed. However, the latter possibility is untenable because X-ray observations show a monotonic increase of the eclipse depth with energy, and in the hard X-ray band ($\sim 60$~keV)  the X-ray eclipse depth exceeds 80\% \citep{2005A&A...437..561C,2013MNRAS.436.2004C}. This suggests that irrespective of the form of the orbit, the optical star edge eclipses the central, hottest parts of the accretion disc in the middle of the X-ray eclipse. Thus, there is an additional argument suggesting that in the model of the rotational broadening of stationary hydrogen emission lines, the broad line wings should disappear in the middle of the eclipse. As this is not observed, there are grounds to reject the model of the double-peak line formation in the outer parts of the accretion disc.

A much more preferable seems to be the model that the Brackett hydrogen emission lines observed by \cite{2017ApJ...841...79R} arise in a circumbinary shell and have composite profiles: the central double-peak component is related to the circumbinary shell rotation and wide wings of these lines are produced in the radial stellar wind outflow from the supercritical accretion disc with a velocity of $\sim 2000$ ~km s$^{-1}$. As the stellar wind forms an extended region, the line formation region in the wind is not fully eclipsed by the optical star, and the wide wings of the Brackett hydrogen lines should not disappear during the eclipses. In this case, the double-peaked structure of the Brackett lines observed by \cite{2017ApJ...841...79R} is formed in an extended rotating circumbinary shell, and the estimates made by \cite{2010A&A...521A..81B,2011A&A...531A.107B,2001ApJ...562L..79B,1988AJ.....96..242F,2010MNRAS.408....2P} that the total mass of SS433 exceeds $\sim 40 M_\odot$ acquire additional support. 

Below we show that the analysis of the long-term
stability of the orbital binary period in SS433 independently confirms this conclusion.

\section{The mass ratio constraints from the stability of the orbital period of SS433}

SS433 demonstrates a surprisingly stable orbital period: during almost 40 years of observations, its period remains constant. According to the recent analysis of SS433 observations over a period of $\sim 28$~years carried out by V.P. Goranskij \citep{2011PZ.....31....5G}, the ephemeris of the primary eclipse minima in SS433 reads:
\beq{e:ephem}
\mathrm{Min\, I}=\mathrm{JD}_{hel}\,2450023.746\pm 0^\mathrm{d}.030+(13^\mathrm{d}.08223\pm 0^\mathrm{d}.00007)\cdot E\,.
\eeq

By assuming a three-sigma rms error of the orbital period  $0^\mathrm{d}.00021$ to be an upper limit on the possible orbital period change over  $\sim 28$~years, it is possible to constrain the mass ratio $q$. To do this, we will use a model of the mass loss from SS433 due to isotropic re-emission with additional taking into account the evidence for the presence of a circumbinary envelope in SS433 which is formed because of the mass loss through the outer Lagrangian point L$_2$.

\begin{figure}
	\includegraphics[width=\columnwidth]{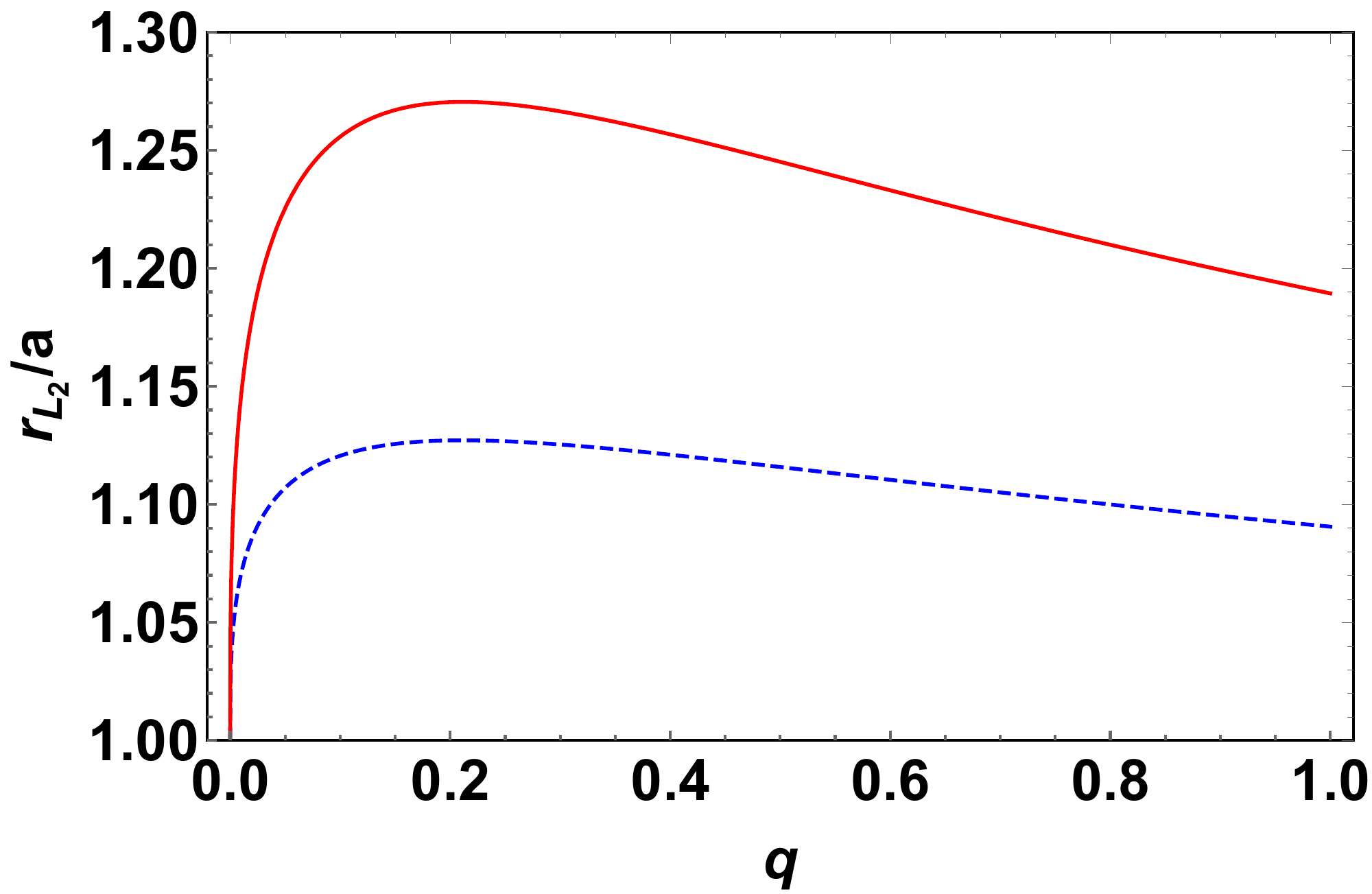}
    \caption{Distance to the 
    L$_2$ point from the binary system's barycenter $r_{L_2}$ in units of the relative orbital separation $a$ for mass ratio $0\le q\le 1$ (the solid upper curve, \Eq{e:rL3}). The dashed curve shows the value $\sqrt{r_{L_2}/a}$ determining the angular momentum loss through L$_2$.}
    \label{f:L2}
\end{figure}
\begin{figure*}
	\includegraphics[width=\columnwidth]{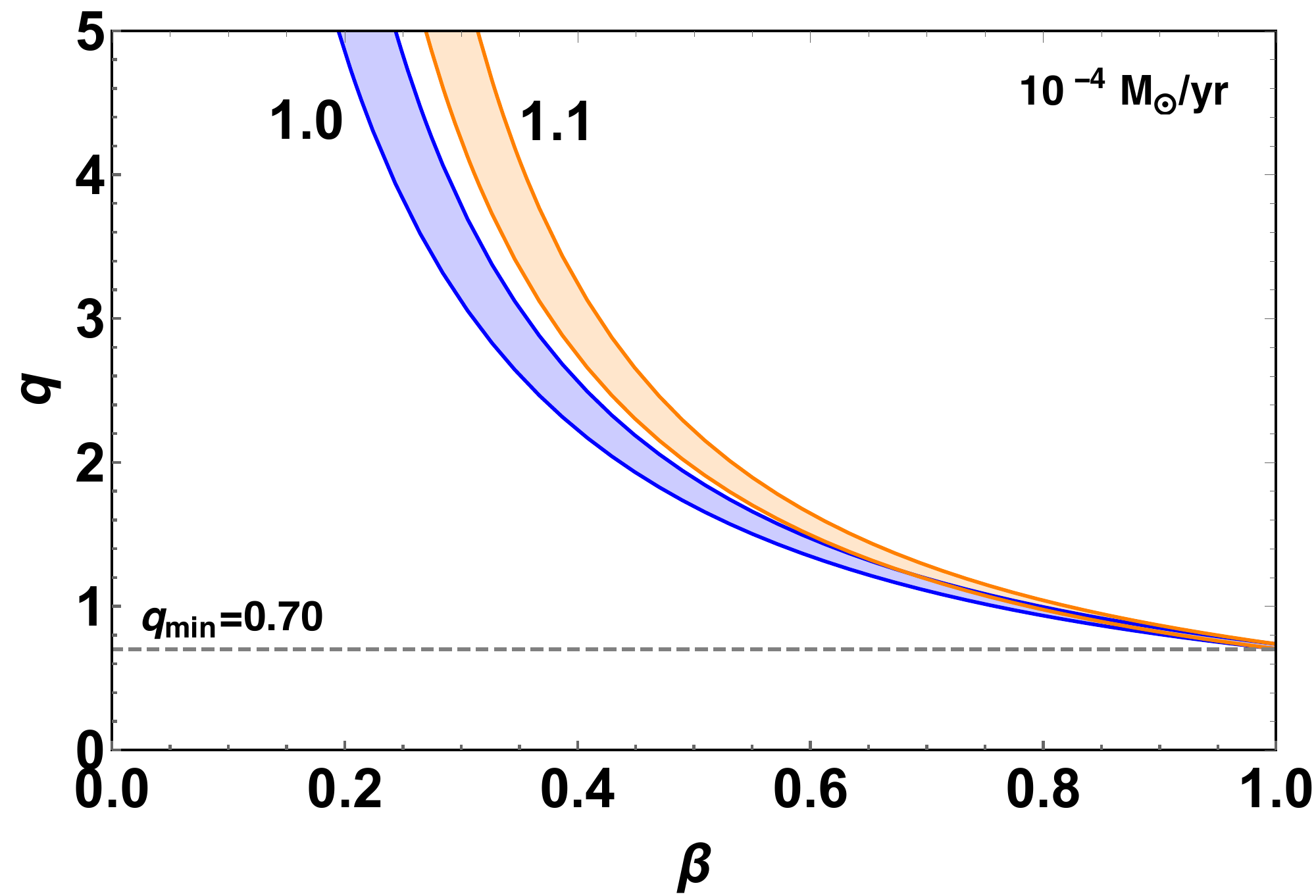}\hfill
    \includegraphics[width=\columnwidth]{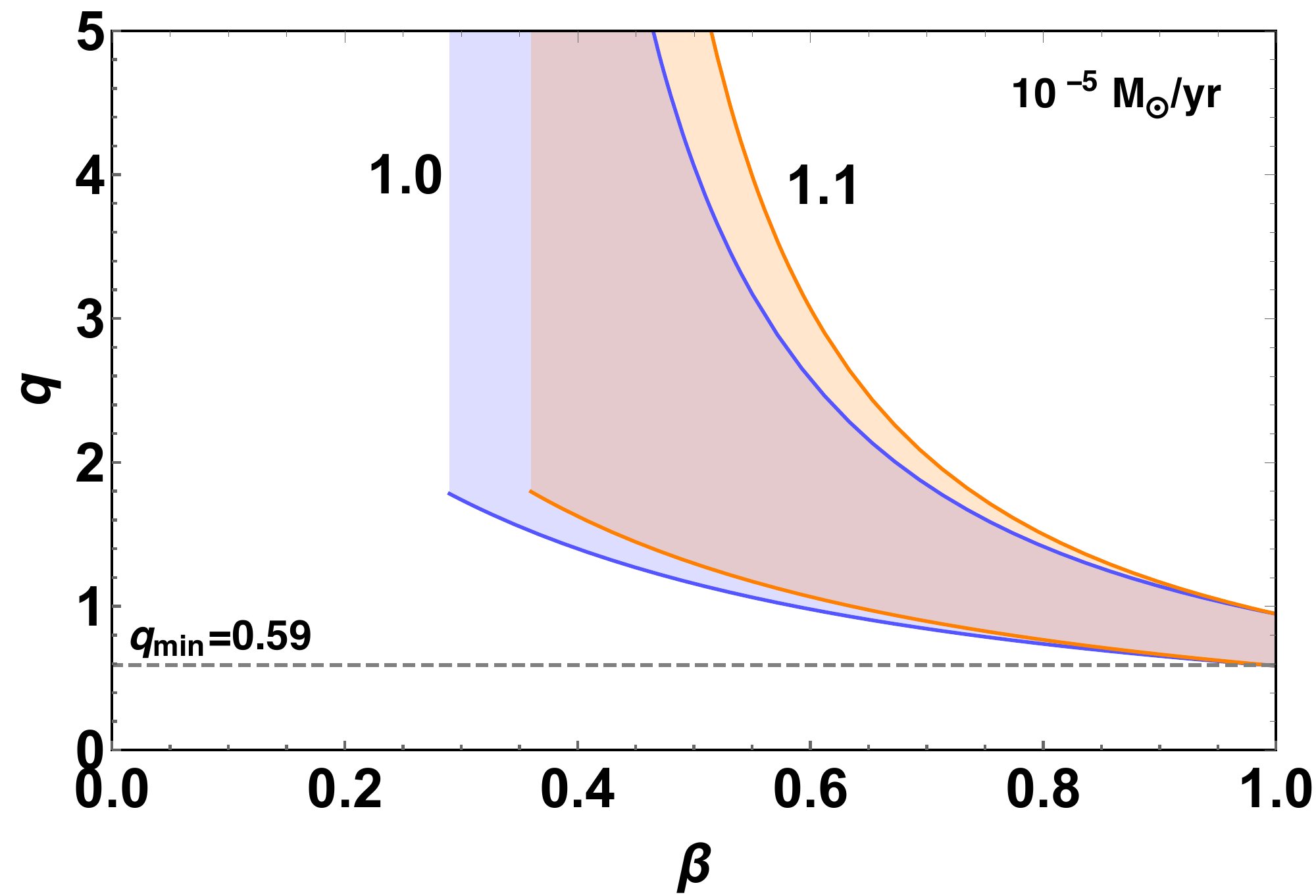}
    \caption{Constraints on the mass ratio of SS433 for different values of 
    parameters $x=1.0, 1.1$ and $\alpha=0.086$ (left panel) and 0.86 
    (right panel) corresponding to the mass-loss rate $\dot M_\mathrm{v}=10^{-4}$ and $10^{-5} M_\odot$ yr$^{-1}$, respectively. }
    \label{f:q(beta)}
\end{figure*}

The model can be specified as follows. Consider a binary system with the component masses $M_\mathrm{v}$ and $M_\mathrm{x}=qM_\mathrm{v}$, the mass ratio $q\le 1$ and the relative orbital separation $a$. The total mass of the system is $M=M_\mathrm{v}+M_\mathrm{x}=M_\mathrm{v}(1+q)$. For simplicity we will consider a circular orbit, which is justified by the high mass transfer rate in the binary. Assume a stable mass loss rate from the optical star $\dot M_\mathrm{v}<0$ and the mass growth rate of the accretor is $\dot M_\mathrm{x}=(\eta-1) \dot M_\mathrm{v}\ge 0$, $0\le \eta\le 1$.  In this case, the conservation of angular momentum implies (see, e.g., \citealt{1997A&A...327..620S,2006LRR.....9....6P})
\beq{e:adota1}
\frac{\dot a}{a}=-2\left(1+(\eta-1)\frac{M_\mathrm{v}}{M_\mathrm{x}}-\frac{\eta}{2}\frac{M_\mathrm{v}}{M}\right)\frac{\dot M_\mathrm{v}}{M_\mathrm{v}}+2\frac{dJ/dt}{J}\,,
\eeq
where $J$ is the orbital angular momentum of the binary.
Conservative mass transfer corresponds to $\eta=0$ and $dJ=0$, and in the non-conservative case $(dJ/dt)/J$ specifies the orbital angular momentum loss by the mass escaping the binary. 

In the isotropic re-emission mode, the mass is transferred to the accretor $M_\mathrm{x}$ and then spherically-symmetrically expelled carrying out a specific angular momentum of the  accretor. For SS433 $\eta\approx 1$ is a good approximation (most of the accreted matter is expelled by the supercritical accretion disc). In this mode, the angular momentum loss term reads: 
\beq{}
\left.\frac{dJ/dt}{J}\right|_\mathrm{iso}=\beta \frac{\dot M_\mathrm{v}}{M}
\frac{M_\mathrm{v}}{M_\mathrm{x}}
\eeq
Here $0\le\beta\le 1$ characterizes the fraction of mass-loss 
in the isotropic re-emission mode. 

We will also assume that a fraction $(1-\beta)$ of the mass loss occurs through the 
outer Lagrangian point L$_2$ behind the less massive component (accretor) forming a circumbinary ring. In this case (see, e.g. \cite{1997A&A...327..620S}), the angular momentum from the system is lost through the L$_2$ point and 
\beq{}
\left.\frac{dJ/dt}{J}\right|_\mathrm{cbr}=(1-\beta) \frac{\dot M_\mathrm{v}}{M}
\frac{M^2}{M_\mathrm{x}M_\mathrm{v}}\sqrt{\frac{a_\mathrm{cbr}}{a}}
\eeq
where $a_\mathrm{cbr}$ is the circumbinary Keplerian ring radius. In the first approximation, we can assume $a_\mathrm{cbr}=r_{L_2}$, 
where the distance from system's barycenter to $L_2$ point 
can be written as a power series in the modified mass ratio $M_x/M=q/(1+q)$
(see, e.g., \cite{1965Icar....4..273D,1983SvA....27..442E}):
\begin{eqnarray}
\label{e:rL3}
&\frac{r_{L_2}}{a}\approx \frac{1}{1+q}+\nonumber \\
&+\myfrac{q}{3(1+q)}^{1/3}+\frac{1}{3}\myfrac{q}{3(1+q)}^{2/3}-\frac{1}{9}\myfrac{q}{3(1+q)}+\frac{50}{81}\myfrac{q}{3(1+q)}^{4/3}
\end{eqnarray}
In the mass ratio range $0\le q\le 1$ this distance falls within the range from $1$ to $\simeq 1.27$ (Fig. \ref{f:L2}).

Summarizing, in the case $\eta=1$ (no accretor mass growth) we find
\beq{e:adota2}
\frac{\dot a}{a}=-2\left(1-\frac{1}{2}\frac{M_\mathrm{v}}{M}\right)\frac{\dot M_\mathrm{v}}{M_\mathrm{v}}+2\beta\frac{M_\mathrm{v}}{M_\mathrm{x}}\frac{\dot M_\mathrm{v}}{M}+2(1-\beta)x\frac{M^2}{M_\mathrm{x}M_\mathrm{v}}\frac{\dot M_\mathrm{v}}{M}\,,
\eeq
where $x\equiv\sqrt{r_{L_2}/a}$. 
Differentiating the third Kepler's law,
$2\dot P_b/P_b=3(\dot a/a)-\dot M/M$, and noticing that $\dot M/M=\dot M_\mathrm{v}/M$ ($\eta=1$), 
we can recast
this equation to the form:
\beq{e:PdotP}
\frac{\dot P_b}{P_b}=-\frac{\dot M_\mathrm{v}}{M_\mathrm{v}}\frac{3(1+2q)q-6\beta-6x(1-\beta)(1+q)^2+q}{2q(1+q)}\,.
\eeq
Therefore, the observational constraint $|\dot  P_b/P_b|<\epsilon$ can be written in the form of two inequalities:
\beq{e:mod+}
A_-(x,\alpha,\beta)q^2+B_-(x,\alpha,\beta)q+C(x,\alpha)\le 0
\eeq
\beq{e:mod-}
A_+(x,\alpha,\beta)q^2+B_+(x,\alpha,\beta)q+C(x,\alpha)\ge 0
\eeq
with coefficients:
\begin{eqnarray}
\label{e:coef}
&A_-(x,\alpha,\beta)=1-\frac{1}{3}\alpha-x(1-\beta)\,\\
&B_-(x,\alpha,\beta)=\frac{2}{3}(1-\frac{1}{2}\alpha-3x(1-\beta))\,\\
&A_+(x,\alpha,\beta)=1+\frac{1}{3}\alpha-x(1-\beta)\,\\
&B_+(x,\alpha,\beta)=\frac{2}{3}(1+\frac{1}{2}\alpha-3x(1-\beta))\,\\
&C(x,\alpha)=-\beta-x(1-\beta)\,.
\end{eqnarray}
Here $\alpha= \epsilon \left|\frac{M_\mathrm{v}}{\dot M_\mathrm{v}}\right|$
is the dimensionless coefficient that can be estimated from observations. Taking the uncertainty in the period ephemeris $\sigma=0^\mathrm{d}.00007$ obtained over $\sim 28$ years of observations \citep{2011PZ.....31....5G} and assuming the optical star mass $M_\mathrm{v}=15 M_\odot$ with the mass-loss rate $\dot M_\mathrm{v}=10^{-4} M_\odot$~yr$^{-1}$, we get the estimate:
\beq{e:alpha}
\alpha=\frac{3\sigma}{\Delta t}\frac{M_\mathrm{v}}{\dot M_\mathrm{v}}= \frac{0^\mathrm{d}.00021}{13^\mathrm{d}.082\times 28\mathrm{yrs}}\frac{15 M_\odot}{10^{-4}M_\odot\mathrm{yr}^{-1}}\approx 0.086
\eeq
Note here that in this estimate the precise value of the optical star mass is rather unimportant because the mass-loss rate is much more uncertain.
The limitation on the mass ratio $q$ can be readily found from inequalities \eqn{e:mod+}, \eqn{e:mod-} and are shown in Fig. \ref{f:q(beta)} for two values of the total mass-loss rate from the system: $\dot M=10^{-4} M_\odot$~yr$^{-1}$ and $10^{-5} M_\odot$~yr$^{-1}$ (left and right panels, respectively). 
At given $\beta$, the interval of $q$ for each value of the parameter $\beta$ corresponds to the uncertainty $\alpha$ in the orbital period change measurement \eqn{e:alpha}.
It is seen that in the limit $\beta\to 1$ (i.e. in the absence of the mass-loss rate through the L$_2$ point), there is a lower limit on the mass ratio, $q_\mathrm{min}\approx 0.7$ and $q_\mathrm{min}\approx 0.6$ for the two assumed mass-loss rates, respectively. In this limit, the inequalities \eqn{e:mod+} and \eqn{e:mod-} yield the allowed intervals of the mass ratio $0.7\lesssim q\lesssim 0.74$ and $0.59\lesssim q\lesssim 0.94$ for 
$10^{-4}$ and $10^{-5} M_\odot$~yr$^{-1}$, respectively. 

\section{Discussion and conclusions}

In this paper, we have analyzed different indirect estimates of the mass ratio and the total mass in SS433. Due to a huge optical luminosity of the supercritical accretion disc and powerful radiation-driven wind outflow, dynamical estimates of the mass ratio of the binary components cannot be fully reliable. 

We note that if the possible overfilling of the Roche lobe by the optical star is taken into account, the analysis of the primary X-ray eclipse both in the standard (2-10 keV) and hard (20-60 keV) X-ray bands suggests a high mass ratio $q=M_\mathrm{x}/M_\mathrm{v}\gtrsim 0.3$. 

We show that the stationary double-peak hydrogen Brackett emission lines observed by \cite{2017ApJ...841...79R} are unlikely to be formed in the outer parts of the accretion disc around the compact star and should be produced in a circumbinary envelope. The existence of such an envelope in SS433 has been repeatedly put forward from radio, optical and IR observations 
\citep{2001ApJ...562L..79B,2010A&A...521A..81B,2011A&A...531A.107B,2011A&A...534A.112B,2010MNRAS.408....2P}. In this model, the total mass of the components in SS433 $M_\mathrm{v}+M_\mathrm{v}\gtrsim 40 M_\odot$, given the radius of the stationary circumbinary disc being at $\sim 1.5a$ ($a$ is the binary orbit separation) see, e.g., Table 1 in \cite{2010A&A...521A..81B}. The smaller radius of the disc would decrease the total mass estimate. However, in this case, a significant variability of the double-peak hydrogen emissions would be expected caused by non-stationary gas flow in the circumbinary shell immediately close to the L$_2$ point, which, apparently, is not observed.

We assume that the emission hydrogen line profiles have double structure for two reasons (see Section 3): the central double-peak component is produced by the rotating circumbinary shell, and the wide wings of these lines are formed in a strong wind outflow from the bright supercritical accretion disc around the compact object.

We have also analyzed the mass loss from the system in the model of isotropic re-emission mode together with a mass loss from the external Lagrangian point L$_2$. In this model, the observed long-term stability of the binary orbital period of SS433 over $\sim 28$ years sets a lower limit on the mass ratio $q\gtrsim 0.6$.

The dynamical analysis given above assumed a particular model for mass loss: an isotropic re-emission stellar wind from the supercritical accretion disc and possible mass-loss rate though the outer Lagrangian point L$_2$ with the formation of a circumbinary ring: $\dot M= \dot M_\mathrm{v}=\beta\dot M_\mathrm{v}|_\mathrm{iso} +(1-\beta)\dot M_\mathrm{v}|_\mathrm{cbr}$. 

We have used the usual assumption that the specific angular momentum carried out by the disc wind is equal to the specific orbital angular momentum of the relativistic component. This is justified in so far as the supercritical disc spherization radius \citep{1973A&A....24..337S} is well within the Roche lobe of $M_\mathrm{x}$: $R_\mathrm{sph}=3R_g(\dot M/\dot M_\mathrm{Edd})\simeq 10^{10}(\mathrm{cm})\dot M/(10^{-4} M_\odot \mathrm{yr}^{-1})\ll R_L(M_\mathrm{x})$ (here $R_g\approx 3\times 10^5 (M_\mathrm{x}/M_\odot)$~cm is the Schwarzschild radius of the relativistic star, $\dot M_\mathrm{Edd}\approx 10^{-8}M_\odot (M_\mathrm{x}/M_\odot)$~yr$^{-1}$ is the critical mass rate corresponding to the Eddington luminosity), which is justified for SS433. 

Much more uncertain is the specific angular momentum loss from the L$_2$ point. We parametrize it with the parameter $x=\sqrt{a_\mathrm{cbr}/a}$, assuming a Keplerian circumbinary ring at distance $a_\mathrm{cbr}$ from the system's barycenter. This parameter can be equivalently written as $x=v_{x}/v_\mathrm{cbr}(M/M_\mathrm{v})= v_{x}/v_\mathrm{cbr}(1+q)$, where $v_{x}/v_\mathrm{cbr}$ is the ratio of the orbital velocity of the relativistic star to the Keplerian velocity at the ring distance. Infrared observations of the Paschen and Brackett hydrogen emission line splitting $\sim 250$~km s$^{-1}$ \citep{2017ApJ...841...79R} can be interpreted in terms of the emission from the circumbinary ring, thus $v_\mathrm{cbr}\sim 250$ km s$^{-1}$. The spectroscopy of HeII emission lines \citep{2004ApJ...615..422H} suggests $v_x=168\pm 18$~km s$^{-1}$. Therefore, the expected value of $x$ falls within the range $\sim 1.16-1.36$ for mass ratios $q$ lying in the interval  $0.7-1$. 
 This estimate shows that the interpretation of the observations does not contradict to the presence of a circumbinary disc not far beyond the L$_2$ point in SS433.
 
 Thus we conclude that the estimation of the component mass ratio in SS433 by independent methods yields $q\gtrsim 0.3\div0.6$, and the total mass of the system is most likely $\gtrsim 40 M_\odot$, suggesting that a stellar-mass black hole powers this unique Galactic microquasar.
 
\section*{Acknowledgements}
We thank the anonymous referee for careful reading of the paper and useful remarks.
The authors thank Dr. N.V. Emelyanov for discussions. 
The work of AMCh and AAB is supported by the RSF grant 17-12-01241 (analysis of constraints from X-ray eclipses and circumbinary shell).
The work of KAP (analysis of constraints from mass-loss modes) is supported by the RSF grant 16-12-10519.




\bibliographystyle{mnras}
\bibliography{ss433} 


\bsp	
\label{lastpage}
\end{document}